\documentclass[11pt,aasms4]{aastex}

\usepackage[dvips]{epsfig}

\begin{document}

\title{The Polarization Variability in the Optical Afterglow of GRB~990712}

\author{Gunnlaugur~Bj\"ornsson
and Elina~J.\ Lindfors\\
Science Institute, University of Iceland\\ 
Dunhaga~3, IS--107 Reykjavik, Iceland\\
e-mail:gulli@raunvis.hi.is}

\begin{abstract}
In a recent paper, Rol and colleagues present evidence for a variable 
polarization in the optical afterglow following the gamma-ray burst
GRB~990712. The variation is highly significant, but the position angle 
appears time independent. Contrary to their conclusion, we point out that 
this can in fact be explained with existing afterglow models, namely that 
of a laterally expanding jet.
\end{abstract}



\section{Introduction}
\label{sec:intro}

It is generally accepted that the optical emission from gamma-ray burst 
afterglows is synchrotron radiation from relativistic electrons (e.g.\
M\'esz\'aros \& Rees 1997). 
Models of optical afterglows based on synchrotron emission, and either 
spherical or collimated outflow geometry have successfully been applied to 
a number of sources (e.g.\ Galama et~al.\ 1998; Bloom et~al.\ 1998; 
Holland et~al.\ 2000). As synchrotron radiation under favorable 
conditions can be up to 70\% polarized, polarization measurements of 
optical afterglows have recently been added to the tool box of 
afterglow researchers. 

The first attempt, by Hjorth et~al.\ (1999), resulted in an upper limit  
of 2.3\% for the polarization of GRB~990123 about 18.3 h after the burst.
The polarization level of GRB~990510 was successfully measured by 
Covino et~al.\ (1999) about 18.5 h after the burst, and Wijers et~al.\ (1999) 
about 2 h later. These latter measurements were obtained using the same 
instruments on the same telescope and the polarization remained constant 
at 1.7\% during the 2 hour interval. Wijers et~al.\ (1999) obtained an 
additional measurement at burst age of  43.3 h, but the polarization level 
at that time was marginally detectable at similar level, mainly due to the 
faintness of the source and worse observing conditions.

In a recent preprint, Rol et~al.\ (2000), present polarization measurements 
of GRB~990712 at three different burst ages, 10.6, 16.7 and 34.7 h after the 
gamma ray event. The polarization level varied between the three measurements
from $2.9\%\pm 0.4$\% to $1.2\%\pm 0.4$\% and $2.2\%\pm 0.7$\%, respectively.
An interesting part of the result is that the position angle does {\em not} 
seem to vary over the 24 h period from the first to the last data point. 
Rol et~al.\ (2000) conclude, based on the constant position angle they find, 
that none of the existing models can successfully explain their result. 
The purpose of this {\em Letter} is to point out that it is in fact possible 
to obtain varying degree of polarization {\em and} a constant position angle 
in beamed models.

\section{Variable Polarization From a Collimated Outflow}
\label{sec:obs}

Several models have been put forward to explain how a polarized emission 
may arise in an optical afterglow, despite the fact that the magnetic field 
generated is expected to be highly tangled with no preferred direction and
therefore no net polarization. Examples include the spherically symmetric model 
of Gruzinov \& Waxman (1999), and the polarization scintillation model of 
Medvedev \& Loeb (1999). Recently, Sari (1999; hereafter referred to as S99) 
and Ghisellini \& Lazzati (1999; hereafter GL99), independently and 
simultaneously, showed that a non-zero and variable polarization can arise 
from an almost totally tangled magnetic field that has some degree of
alignment, if the emission arises in a collimated outflow, and the observers line 
of sight is located off the outflow axis but within the collimated beam. 
This polarization variability is essentially a geometrical effect, the most 
important points being the following (we will assume here that all geometries 
are conical and we refer the reader to Fig.\ 2 in GL99 and Figs.\ 2 and 3 in S99): 
Initially, when the expansion is highly relativistic, the observer only receives 
radiation from within the relativistic cone of angular size $1/\Gamma$, where 
$\Gamma$ is the bulk Lorentz factor. This cone is centered on, and symmetric with 
respect to, the line of sight and therefore no polarization is observed. As the 
expansion slows down, the edge of the relativistic cone reaches the edge of the 
collimated beam and thereafter looks asymmetric to the observer. A net polarization 
arises, that reaches a maximum as the relativistic cone expands and looks more and 
more asymmetric, and drops to zero again when the emitting areas contributing to 
the polarization in two different directions (``vertical and horizontal'') 
become equal. The polarization then rises again with increasing asymmetry 
between the areas emitting the two possible polarization directions, but with 
the position angle rotated by $90^\circ$. The polarization finally drops to 
zero again when $\Gamma \rightarrow 1$ (GL99), or exhibits a third 
maximum if the jet is spreading (S99).

In Figure 1a we show a typical evolution of the polarization for a conical 
beam of fixed opening angle, $\theta_c=5^\circ$, with the observers line of 
sight making an angle $\theta_0=f\theta_c$, ($f<1)$, with the cone symmetry 
axis. We used the approach of GL99 to construct the figure and therefore only 
exhibit two maxima. The evolution is shown as a function of the inverse bulk 
Lorentz factor for two different values of $f$. The Lorentz factor can be 
converted to time, using the relation $\Gamma=\Gamma_0 (t/t_0)^{-3/8}$.
For this figure we have used $\Gamma_0=100$ for the initial value of the 
Lorentz factor and $t_0=50$ s. Note that lowering $f$ for a fixed $\theta_c$, 
brings the observers line of sight closer to the symmetry axis and therefore 
decreases the net polarization. It also shifts the first maximum and the 
minimum to later times (lower $\Gamma$), while the second maximum occurs 
almost at the same time ($\Gamma\approx 5$). Note also, that the first maximum 
occurs typically less than an hour after the burst, the minimum less than 10 h 
after the burst and the second maximum from 1-2 days after the burst. 
The polarization level is modestly affected by the radiation spectral index. 
We assume a power law spectral distribution, and take the spectral index 
to be $\beta=0.6$ as e.g.\ observed for GRB~990712 (Sahu et~al.\ 2000).

A crucial effect that is not discussed in detail by GL99 and for which S99
considers only one particular example in his toy model, is the {\em evolution} 
of $f$, the ratio of the angle the line of sight makes with the jet axis to 
that of the collimated beam. The jet axis is most likely defined by the angular 
momentum of the burst progenitor system, and is presumably fixed in space. 
The angle between the jet axis and the line of sight should therefore be 
constant, unless the jet is precessing for which there is no evidence. 
If the jet is expanding laterally, the ratio 
$f=\theta_0/\theta_c$, {\em decreases} with time. Generating a sequence of 
polarization curves as in Fig.\ 1a, varying (increasing) only the jet opening 
angle, shows a decreasing magnitude of both maxima and a shift of the first 
maximum and the minimum to the right (towards lower $\Gamma$ or later times). 
The evolution of the second maximum is particularly interesting as it takes 
place entirely under the polarization curve defined by the initial value of $f$, 
with the maximum occurring at almost constant value of $\Gamma$ 
(see also S99 and GL99). 

We show an example in Figure 1b, with the data points of GRB~990712 superimposed. 
As the first data point is obtained about 11 h after the burst, we assume that 
the polarization at that time has already evolved into the region of second 
maximum and therefore that the position angle has already changed by 90$^\circ$. 
With $\Gamma_0=100$, and $t_0=50$~s, we find that an opening angle of 
$\theta_c=5.1^\circ\pm 0.1^\circ$, with $f=0.9$, fits the first point. We then
let $\theta_c$ increase to $6.0^\circ\pm 0.2^\circ$, over the next 6 h, giving 
$f=0.77$, and finally a modest increase to $6.2^\circ\pm 0.5^\circ$ fits the
last point 18 h later. The last two data points are also consistent with being
on the same polarization light curve. In that case we would be observing a
widening of the collimation angle by $1^\circ$ over a 6 h period between
the first and the second point, and approximately a constant opening angle
thereafter, or a slower rate of lateral expansion that may be due to density 
irregularities in the local environment. We emphasize that the 
above is obtained by taking ``snapshots'' of the evolving polarization light 
curve, where only the jet opening angle has been changed between each shot. 
The ``error estimates'' on the opening angle are determined by searching for 
values of $\theta_c$ that give polarization within the error of the measured 
polarization points at the appropriate time. It is interesting that a modest 
variation in the jet opening angle (about 20\%), can easily change the 
polarization by a factor of two to three. An important consequence of the 
variable polarization being due to temporal evolution of the second maximum 
is a constant position angle, naturally explaining the observations of the 
GRB~990712 afterglow. 

The above analysis is a simple extension of the GL99 model, and complements the 
approach of S99 that assumed that the jet opening angle evolved as $1/\Gamma$, 
once $\Gamma$ had decreased below the inverse of the  initial jet opening angle. 
The initial polarization evolution, i.e.\ the first maximum and minimum, is 
therefore similar in both approaches, differences arising when the jet opening 
angle starts spreading, but the polarization light curve at that time is already 
in the region of second maximum. Despite differences in details of GL99 and S99, 
the location of the the second maximum in both cases occurs on similar time scales, 
about 1-2 days after the burst. We have shown here that a modest variation in the
jet opening angle is sufficient to explain the polarization measurements of
Rol et~al.\ (2000). Numerical simulations using more realistic models are needed 
to follow the detailed temporal evolution of the polarization light curve, 
in particular the evolution of the second maximum.

The optical light curve of GRB~990712 decayed as a power law with an index of
$\alpha\approx-1.0$ (Sahu et~al.\ 2000, Hjorth et~al.\ 2000). This is similar 
to the decay index of GRB~990123 and somewhat steeper than that of GRB~990510, 
before the break in their light curves. In the latter two cases the light curve 
steepened about 1-2 days after the burst (e.g.\ Kulkarni et~al.\ 1999; 
Harrison et~al.\ 1999; Stanek et~al.\ 1999; Israel et~al.\ 1999; Holland et~al.\ 2000). 
A model of a collimated outflow predicts a steepening of the light curve when 
$1/\Gamma \approx\theta_c$, the sharpness of the break depending on the rate 
of lateral expansion (e.g.\ Rhoads 1999), and the break in the light curves of 
GRB~990123 and GRB~990510 indicates a jet opening angle of about $5^\circ$.
Interpreting the polarization data for GRB~990712 with a spreading jet, therefore
implies that the optical light curve should show a break after about 1-2 days. 
The modest increase in the opening angle implied by the polarization measurements
requires the break to have been rather sharply defined in time.
No such break has been reported (Sahu et~al.\ 2000; Hjorth et~al.\ 2000), perhaps 
because the host galaxy is bright and was already affecting the magnitudes of the 
optical transient 10 h after the burst (second data point on the optical light curve).

The polarization measurement of GRB~990123 at 18.3 h and of GRB~990510 at 
18.5 h, 20.7 h and 43.3 h, are likely to have 
been taken during the second maximum. It is crucial that polarization
measurements be attempted as soon as possible after the discovery of an
optical afterglow. In particular, to demonstrate the 90$^\circ$ change in the
position angle, a positive detection well before a burst age of 10 h is needed 
and would nicely confirm the applicability of collimated models. A well sampled 
polarization light curve is a powerful tool in exploring the properties of 
burst afterglows and their surroundings and can potentially provide more 
detailed information than the optical light curve alone.

\acknowledgments

\acknowledgements
This work was supported by the Icelandic Research Council
and the University of Iceland Research Fund. We thank the
anonymous referee for useful suggestions.



\begin{figure}[t]
\epsscale{0.3}
\centerline{\epsfig{figure=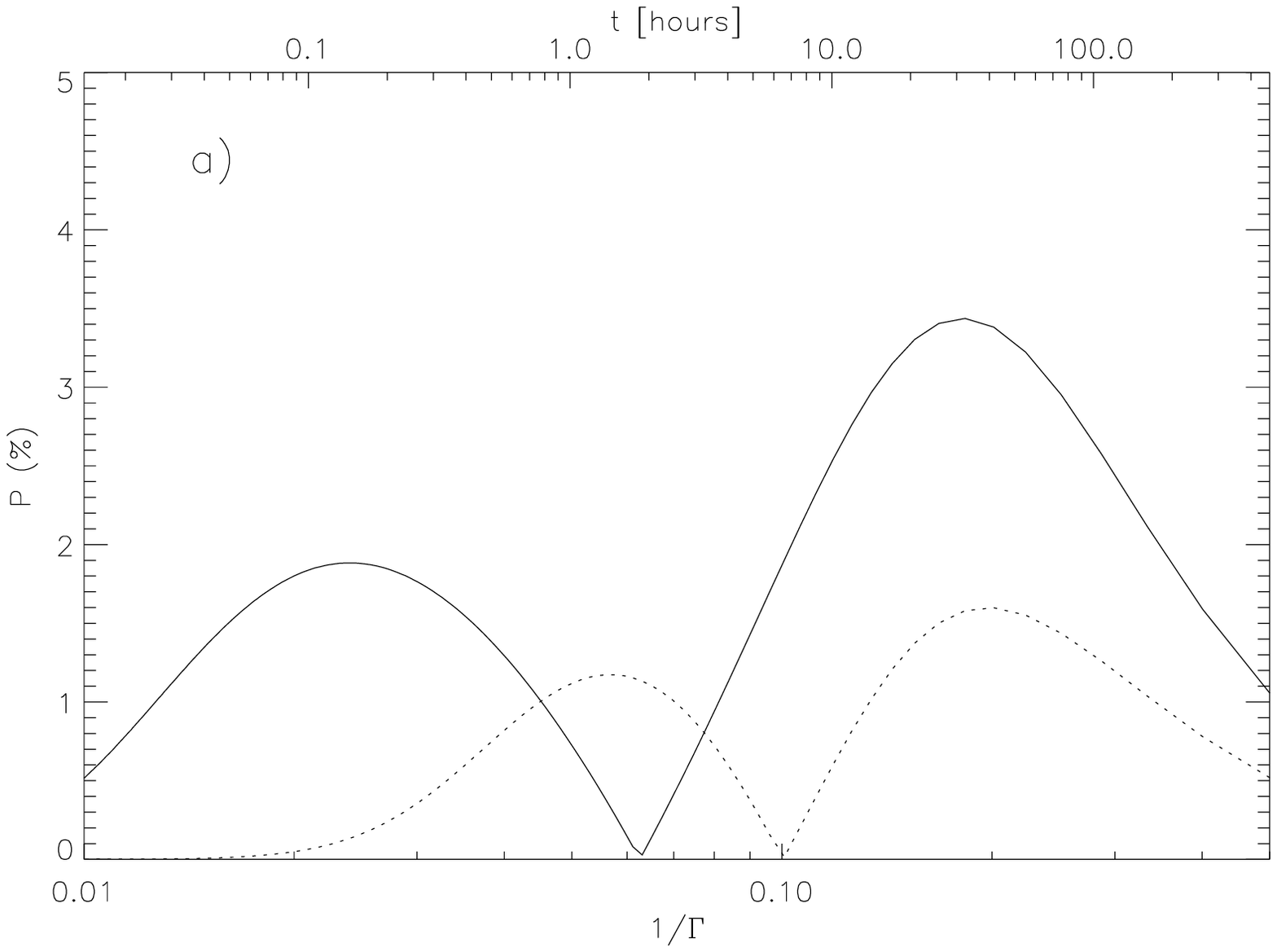,height=6cm,clip=}\epsfig{figure=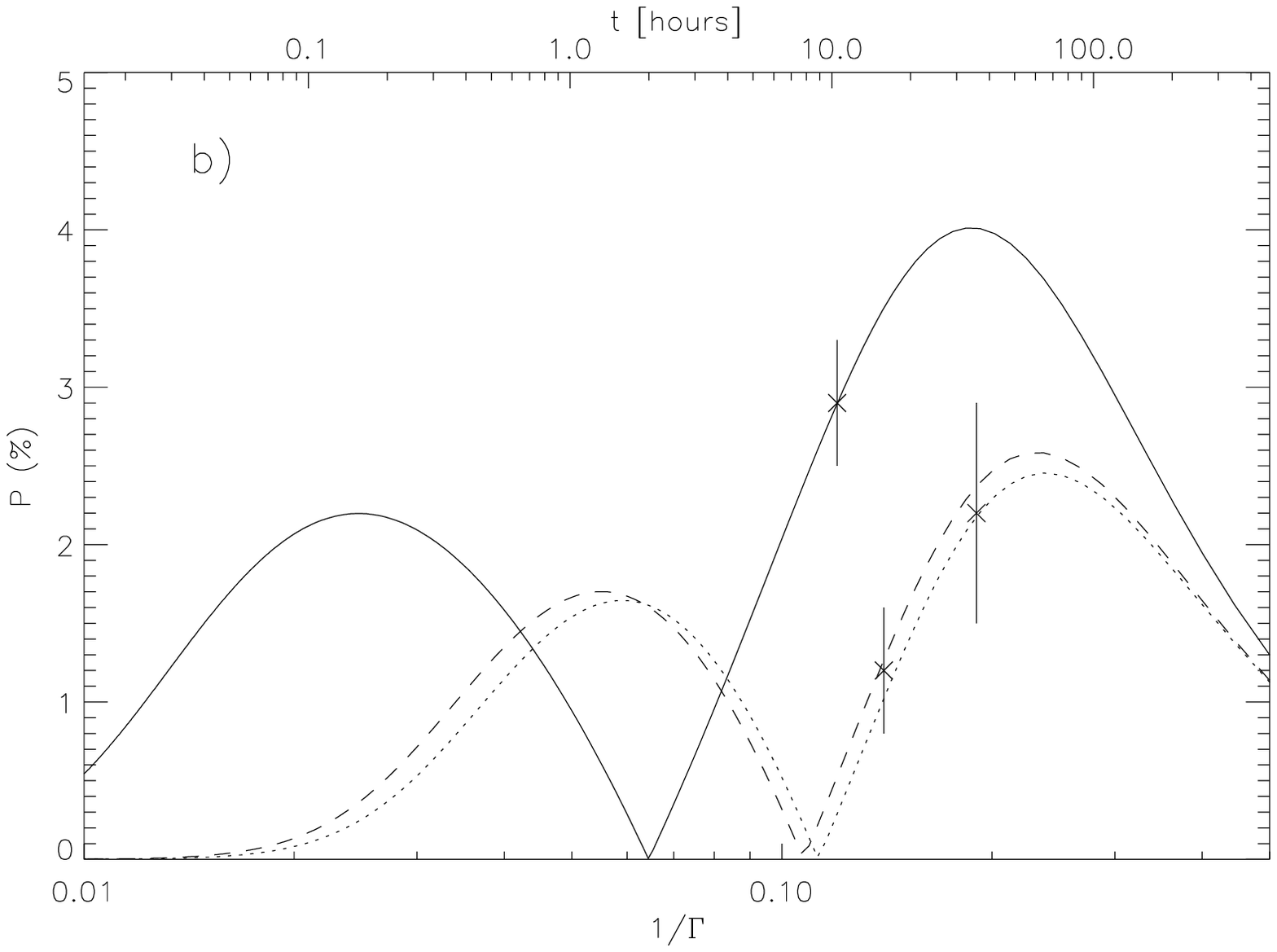,height=6cm,clip=}}
\caption{a) Typical evolution of the polarization light curve for constant 
$\theta_c=5^\circ$, and $f=0.9$ (solid curve) and $f=0.67$ (dotted curve).
Note that the first maximum and the minimum shifts to the right as $f$ is
decreased, while the second maximum only decreases in amplitude. 
b) Polarized light curves of a collimated outflow with a varying opening 
angle and a fixed angle between the jet symmetry axis and the line of
sight. If $\theta_c$ is increased, $f$ decreases. The solid curve has 
$\theta_c=5.1^\circ$ and $f=0.9$, the dashed has $\theta_c=6.0^\circ$ and 
$f=0.77$, and the dotted has $\theta_c=6.2^\circ$ and $f=0.74$. As in panel 
a), the first maximum and the minimum move to the right while the second 
maximum shifts to the right by a factor of 2 in time but does not cross the
initial curve (solid). The polarization values for GRB~990712 
are superimposed. We have assumed that the maximum polarization from 
synchrotron radiation is $P_0=70\%$, in both a) and b).
}
\end{figure}

\end{document}